\documentclass{LMCS}
\usepackage{amsmath, amsthm, amssymb, latexsym, proof}
\usepackage{epic,eepic}
\usepackage{enumerate,hyperref}

\theoremstyle{remark}
\newtheorem*{notation}{Notation}
\newtheorem*{convention}{Convention}

\newcommand{\lp}{\mathcal{L}}
\newcommand{\Ss}{\mathcal{S}}

\newcommand{\T}{\mathcal{T}}
\newcommand{\U}{\mathcal{U}}
\newcommand{\G}{{\bf G}}
\newcommand{\W}{\mathcal{W}}
\newcommand{\F}{\mathcal{F}}

\newcommand{\dd}{\diamond}
\newcommand{\ptil}{p_1 \til p_n}
\newcommand{\psitil}{\psi_1 \til \psi_n}

\newcommand{\affirm}{\vartriangleright}
\newcommand{\waffirm}{\blacktriangleright}

\newcommand{\bi}{\begin{enumerate}[$\bullet$]}
\newcommand{\ei}{\end{enumerate}}
\newcommand{\be}{\begin{enumerate}[(1)]}
\newcommand{\ee}{\end{enumerate}}
\newcommand{\bd}{\begin{enumerate}[\hbox to8 pt{\hfill}]}
\newcommand{\ed}{\end{enumerate}}

\newcommand{\Ra}{\Rightarrow}
\newcommand{\g}{\Gamma}
\newcommand{\de}{\Delta}
\newcommand{\su}{\supset}

\newcommand{\vd}{\vdash}
\newcommand{\vD}{\vDash}
\newcommand{\bs}{\bigskip}
\newcommand{\ms}{\medskip}
\newcommand{\st}{{\ |\ }}   
\newcommand{\til}{,\dots,}

\newcommand{\tup}[1]{\langle #1 \rangle}

\newcommand{\fe}{\varphi}
\newcommand{\suq}{\subseteq}
\newcommand{\dera}[2]{
    \begin{tabular}{c}
    $ #1 $ \\ \hline
    $ #2 $ \\
\end{tabular}}
\newcommand{\derb}[3]{
    \begin{tabular}{c}
    $ #1 \ \ \ \ #2 $ \\ \hline
    $ #3 $ \\
    \end{tabular}}
\newcommand{\derc}[4]{
    \begin{tabular}{c}
    $ #1 \ \ \ \ #2 \ \ \ \ #3 $ \\ \hline
    $ #4 $ \\
    \end{tabular}}

\def\doi{6 (4:12) 2010}
\lmcsheading%
{\doi}
{1--20}
{}
{}
{Jun.~28, 2010}
{Dec.~25, 2010}
{}

\begin{document}

\title[On Constructive Connectives and Systems]
{On Constructive Connectives and Systems}

\author[A.~Avron]{Arnon Avron}	
\address{School of Computer Science, Tel Aviv University, Israel}	
\email{\{aa,orilahav\}@post.tau.ac.il}  

\author[O.~Lahav]{Ori Lahav}	
\address{\vskip-6 pt}	




\begin{abstract}
Canonical inference rules and canonical systems are
defined in the framework of {\em non-strict} single-conclusion sequent 
systems, in which the succeedents of sequents can be empty.
Important properties of this framework are investigated,
and a general non-deterministic Kripke-style semantics is provided.
This general semantics is then used  to
provide a constructive (and very
natural), sufficient and necessary {\em coherence} criterion for
the validity of the strong  cut-elimination theorem in such a system.
These results suggest new syntactic and semantic 
characterizations of basic constructive connectives.
\end{abstract}

\maketitle

\section{Introduction}

There are two traditions concerning the definition and 
characterization of logical connectives. 
The better known one is the semantic tradition, which is
based on the idea that an $n$-ary connective $\diamond$
is defined by the conditions which make
a sentence of the form ${\diamond (\fe_1 \til \fe_n)}$ 
{\em true}. The other is the proof-theoretic tradition (originated
from \cite{gentzen69} ---  see e.g.  \cite{handbook-sundholm} for discussions and references).
This tradition implicitly divides the connectives into {\em basic} connectives and compound connectives,
where the latter are defined in terms of the basic ones. 
The meaning of a basic connective, in turn, is determined by 
a set of derivation rules which are associated with it. 
Here one usually has in mind a natural deduction or a sequent system,
in which every logical rule is an introduction rule
(or perhaps an elimination rule, in the case of natural deduction) 
of some unique connective.
However, it is well-known that not every set of rules can be taken as a definition of a 
basic connective. 
A minimal requirement is that whenever some sentence involving exactly one basic connective
is provable, then it has a proof which involves no other connectives.
In ``normal" sequent systems, in which every rule except cut has the subformula property,
this condition is guaranteed by a cut-elimination theorem. Therefore only sequent systems for which
such a theorem obtains are considered as useful for defining connectives.

In \cite{AL05} the semantic and the proof-theoretic traditions were shown to be equivalent
for a a large family of what may be called semi-classical connectives 
(which includes all the classical connectives, as well as many others).
In these papers multiple-conclusion
canonical (= `ideal') propositional rules and systems were defined in precise terms.
A simple coherence criterion for the non-triviality of such a system was given,
and it was shown that a canonical system 
is coherent if and only if it admits cut-elimination.
Semi-classical connectives were characterized using canonical rules in coherent canonical systems.
In addition, each of these connectives was given  
a semantic characterization. This characterization uses  
two-valued {\em non-deterministic} truth-tables -- a
natural generalization of the classical truth-tables. 
Moreover, it was shown there how to translate a semantic definition of a connective 
to a corresponding proof-theoretic one, and vice-versa.\footnote{It might be interesting to note that 
every connective in this framework can be viewed as basic.}


In this paper we attempt to provide similar 
characterizations for the class of {\em basic constructive connectives}. 

What exactly is a constructive connective? Several different answers
to this question have been given in the literature,
each adopting either of the traditions described above
(but not both!). 
Thus in \cite{McCullough} McCullough
gave a purely semantic characterization of constructive connectives,
using a generalization of the Kripke-style semantics for intuitionistic logic.
On the other hand Bowen suggested in \cite{Bowen} 
a quite natural proof-theoretic criterion for (basic)
constructivity:  an $n$-ary connective $\diamond$, 
defined by a set of sequent rules, is constructive if
whenever a sequent of the form ${\Ra\diamond (\fe_1 \til \fe_n)}$ is provable, 
then it has a proof ending by an application of one 
of right-introduction rules for $\diamond$.



In what follows we generalize and unify the 
syntactic and the semantic approaches by adapting
the ideas and methods used in \cite{AL05}. 
The crucial observation on which our theory is based is that every
connective of a ``normal" {\em single-conclusion} sequent system
that admits cut-elimination is necessarily constructive 
according to Bowen's criterion (because without using
cuts, the only way to derive ${ \Rightarrow \diamond (\fe_1 \til \fe_n)}$
in such a system is to prove first the premises of one of 
its right-introduction rules). 
This indicates that only single-conclusion sequent rules are 
useful for defining constructive connectives.
In addition, for defining {\em basic} connectives, 
only {\em canonical} derivation rules (in a sense similar to that used 
in \cite{AL05}) 
should be used.
Therefore, our proof-theoretic characterization
of {\em basic} constructive connectives is done using 
cut-free single-conclusion canonical systems. 
These systems are the natural constructive counterparts of the 
multiple-conclusion canonical systems of \cite{AL05}.
On the other hand,
McCullough's work suggests that 
an appropriate counterpart of the semantics of non-deterministic truth-tables 
should be given by a non-deterministic generalization 
of Kripke-style semantics.


General single-conclusion canonical rules and systems were first introduced 
and investigated in \cite{AL10}. 
A general non-deterministic Kripke-style semantics for such systems 
was also developed there, and a constructive necessary and
sufficient {\em coherence} criterion 
for their non-triviality was provided. Moreover: it was shown
that a system of this
kind admits a strong form of cut-elimination iff it is coherent. 
However, \cite{AL10} dealt only with {\em strict} single-conclusion systems,
in which the succeedents of sequents contain {\em exactly} one formula.
Unfortunately, in such a framework it is impossible to have canonical rules even 
for a crucial connective like intuitionistic  negation.
To solve this, we move here to Gentzen's original 
(non-strict) single-conclusion framework, 
in which the succeedents of sequents contain {\em at most} one formula. 
There is a price to pay, though, for this extension of the framework. 
As we show below, in this more general framework we lose
the equivalence between the simple coherence criterion
of \cite{AL05,AL10} and non-triviality,
as well as the equivalence proved there between simple cut-elimination and strong cut-elimination.
Hence the theory needs some major changes.


In the rest of this paper 
we first redefine the notions of a canonical inference rule and a canonical
system in the framework of non-strict single-conclusion sequent systems.
Then we turn to the semantic point of view, and 
present a corresponding general non-deterministic Kripke-style semantics.
We show that every canonical system 
induces a class of non-deterministic Kripke-style frames,
for which it is strongly sound and complete. 
This general semantics is then used  to show that a   canonical system
${\bf G}$ is coherent iff it admits a strong form of non-triviality, and this happens
iff the strong  cut-elimination theorem is valid for ${\bf G}$.


Taken together, the results of this paper suggest that a basic constructive connective
is a connective that can be defined using a set of canonical rules
in a coherent (non-strict) single-conclusion sequent system.
We show that this class is broader than that suggested in \cite{McCullough},
and includes connectives that cannot be expressed by the four basic intuitionistic connectives.
Examples include the 
``converse non-implication" and ``not-both" connectives from \cite{Bowen}, 
as well as the weak implication of primal intuitionistic logic from \cite{gu09}.
These connectives were left out by McCullough's deterministic semantic characterization
because their semantics is strictly non-deterministic.

\section{Preliminaries}

In what follows  $\lp$ is a propositional language,
${\F}$ is its set of wffs,
$p,q$ denote atomic formulas,
$\psi,\fe,\theta$ denote arbitrary formulas  (of $\lp$),
$\T$ and $\U$ denote subsets of ${\F}$,
$\g,\de,\Sigma,\Pi$ denote {\em finite} subsets of ${\F}$,
and $E,F$ denote subsets of ${\F}$ with at most one element.
We assume that the atomic formulas of $\lp$
are $p_1,p_2,\ldots$ (in particular:
$\ptil$ are the first $n$ atomic formulas of $\lp$).

\begin{notation}
For convenience we sometimes discard  parentheses for sets,
and write e.g. just $\psi$ instead of $\{\psi\}$.
We also employ other standard abbreviations, like
$\g,\de$ instead of $\g \cup \de$.
\end{notation}

\begin{defi}
\label{tcr}
A {\em Tarskian consequence relation} ({\em tcr} for short) for $\lp$
is a binary relation $\vd$ between sets of formulas of $\lp$ and formulas of $\lp$
that satisfies the following conditions:
\begin{tabbing}
\ \ \ \ \=
 {\it Strong} {\it Reflexivity}: \ \ \ \=
if $\fe \in \T$ then $\T \vd \fe$.\\
 \> {\it Monotonicity}: \> if $\T \vd \fe$ and
$\T \suq \T^\prime$ then $\T^\prime \vd \fe$. \\
\index{cut}
 \> {\it Transitivity} {(\it cut)}:
\> if $\T \vd \psi$ and $\T, \psi \vd \fe$
then $\T \vd \fe$.
\end{tabbing}
\end{defi}

\noindent In the non-strict framework, it is natural to extend Definition \ref{tcr} as follows: 

\begin{defi}
\label{etcr}
An {\em Extended Tarskian consequence relation} ({\em etcr} for short) for $\lp$
is a binary relation $\vd$ between sets of formulas of $\lp$ and singletons or
empty sets of formulas of $\lp$ that satisfies the following conditions:
\begin{tabbing}
\ \ \ \ \=
 {\it Strong} {\it Reflexivity}: \ \ \ \=
if $\fe \in \T$ then $\T \vd \fe$.\\
 \> {\it Monotonicity}: \> if $\T \vd E$, 
$\T \suq \T^\prime$, and $E\suq E^\prime$, then $\T^\prime \vd E^\prime$. \\
\index{cut}
 \> {\it Transitivity} {(\it cut)}:
\> if $\T \vd \psi$ and $\T, \psi \vd E$
then $\T \vd E$.
\end{tabbing}
\end{defi}

\noindent Intuitively, ``$\T \vd\ $" means that $\T$ is inconsistent 
(i.e. $\T \vd \fe$ for every formula $\fe$). 

\begin{defi}
\label{substitution}
An {\em $\lp$-substitution} is a function $\sigma : {\F} \to {\F}$, 
such that for every $n$-ary connective $\dd$ of $\lp$, we have:
$\sigma(\dd(\psitil))=\dd(\sigma(\psi_1) \til \sigma(\psi_n))$.
Obviously, a substitution is determined by the values it assigns to atomic formulas.
A substitution is extended to sets of formulas in the obvious way:
 $\sigma(\T)=\{\sigma(\fe)\st \fe\in\T\}$
(in particular, $\sigma(\emptyset)=\emptyset$).
\end{defi}

\begin{defi}
\label{etcr properties}
An etcr $\vd$ for $\lp$ is {\it structural} 
if for every $\lp$-substitution $\sigma$ and every $\T$ and $E$, if
$\T\vd E$ then $\sigma(\T) \vd \sigma(E)$.
$\vd$ is {\em finitary} iff the following condition
holds for every $\T$ and $E$:
if $\T \vd E$ then there exists a
finite $\g \suq \T$ such that
$\g \vd E$.
$\vd$ is {\em consistent} (or {\em non-trivial})
if $p_1\not\vd p_2$.
\end{defi}

It is easy to see that 
there are exactly {\em four} inconsistent structural etcrs 
in any given language: $\T\vd E$ for every $\T$ and $E$;
$\T\vd E$ for every $E$ and nonempty $\T$;
$\T\vd E$ for every $\T$ and nonempty $E$;
and $\T\vd E$ for every nonempty $\T$ and nonempty $E$.
These etcrs are obviously trivial, so we exclude them from
our definition of an {\em extended logic}:

\begin{defi}
\label{elogic}
A propositional {\it extended logic} is a pair $\tup{\lp,\vd}$,
where $\lp$ is a propositional language, and $\vd$ is
an etcr for $\lp$ which is structural, finitary, and consistent. 
\end{defi}

Sequents, which are the main tool for introducing extended logics, are defined as follows:

\begin{defi} 
A {\em non-strict single-conclusion sequent} is an expression of the form $(\g\Ra E)$ where
$\g$ and $E$ are finite sets of formulas, and $E$ is either a singleton or empty.  
A {\em non-strict single-conclusion Horn clause} is 
a non-strict single-conclusion sequent which consists of atomic formulas only.
\end{defi}

\begin{convention}
From now on, by ``{\em sequent (clause)}" we shall mean 
``non-strict single-conclusion sequent (Horn clause)".
\end{convention}

\begin{defi} 
A sequent of the form $(\g\Ra\{\fe\})$ is called {\em definite}.  
A sequent of the form $(\g\Ra\emptyset)$ is called {\em negative}.
\end{defi}

\begin{notation}
We mainly use $s$ to denote a sequent and $\Ss$ to denote a set of sequents.
We usually omit the outermost parentheses of sequents to improve readability.
For convenience, we shall denote a sequent of the form $\g\Ra\emptyset$ by $\g\Ra\ $,
and a sequent of the form $\g\Ra\{\fe\}$ by $\g\Ra\fe$.
\end{notation}

\section{Canonical Systems}

The following definitions formulate in exact terms the structure of 
sequent rules (and systems) that can be used to define basic constructive connectives.
We first define right-introduction rules and their applications, and then deal with left-introduction rules.

\begin{defi}
\label{canonical right-introduction rule}
\
\be
\item
A {\em  single-conclusion canonical right-introduction rule} for a connective $\diamond$ of arity $n$
is an expression of the form:
\begin{center}
${\{\Pi_i \Ra E_i\}_{1 \leq i \leq m} / \  \Ra \dd(\ptil)}$
\end{center}
\sloppy
\noindent
where
$m\geq 0$, and ${\Pi_i\cup E_i \suq \{\ptil\}}$
for every  $1 \leq i \leq m$.
%
The clauses 
$\Pi_i \Ra E_i$ ($1 \leq i \leq m$) are the {\em premises} of the rule,
while  ${\Ra \dd(\ptil)}$ is its {\em conclusion}. 
\item An {\em application} of the rule
$\{\Pi_i \Ra E_i\}_{1 \leq i \leq m} / \ \Ra \dd(\ptil)$
is any inference step of the form:
\[\dera{\{\g,\sigma(\Pi_i)\Ra \sigma(E_i) \}_{1 \leq i \leq m}}
{\g\Ra \sigma(\dd(\ptil))}\]
where $\g$ is a finite set of formulas and $\sigma$ is an $\lp$- substitution.
\ee
\end{defi}

A canonical right-introduction rule may have negative premises 
(negative sequents serving as premises).
Obviously, in applications of such a rule, a right 
context formula cannot be added
to its negative premises. 
Left-introduction rules are somewhat more complicated, since
in their applications it is not impossible to add a right context formula to 
the negative premises and to the conclusion. However,
in the general case there might also be negative premises 
which do {\em not} allow such an  addition of  a right context.
Accordingly, in what follows we split the set of premises
of a canonical left-introduction rule into two sets: {\em hard} premises 
which do not allow right context, 
and {\em soft} premises, which do allow it.

\sloppy

\begin{defi}
\label{canonical left-introduction rule}
\
\be
\item
A {\em  single-conclusion canonical left-introduction rule} for a connective $\diamond$ of arity $n$
is an expression of the form:
\begin{center}
${\tup{\{\Pi_i \Ra E_i\}_{1 \leq i \leq m},\{\Sigma_j \Ra \}_{1 \leq j \leq k}} / \ \dd(\ptil)\Ra}$
\end{center}
where $m,k\geq 0$, $\Pi_i\cup E_i\suq\{\ptil\}$ for $1 \leq i \leq m$, 
and $\Sigma_j\suq\{\ptil\}$  for $1 \leq j \leq k$. The clauses
$\Pi_i \Ra E_i$ ($1 \leq i \leq m$) are called the {\em hard premises}
of the rule, 
 $\Sigma_j \Ra$ ($1 \leq j \leq k$)  are its {\em soft premises}, 
and $\dd(\ptil) \Ra$ is its {\em conclusion}.
\item An {\em application} of the rule
\\ $\tup{\{\Pi_i \Ra E_i\}_{1 \leq i \leq m},\{\Sigma_i \Ra \}_{1 \leq i \leq k}} / \dd(\ptil)\Ra$
is any inference step of the form:
\[\derb{\{\g,\sigma(\Pi_i) \Ra \sigma(E_i)\}_{1 \leq i \leq m}}
 {\{\g,\sigma(\Sigma_i) \Ra E\}_{1 \leq i \leq k}}
 {\g, \sigma(\dd(\ptil))\Ra E}\]
where $\g\Ra E$ is an arbitrary sequent, and $\sigma$ is an $\lp$- substitution.
\ee
\end{defi}

\fussy

\begin{rem}
Note that definite premises of a canonical left-introduction rules 
are all hard premises, as they do not allow the addition of a right context.
\end{rem}

\begin{convention}
From now on, by  ``{\em canonical right-introduction (left-introduction) rule}" 
we shall mean ``single-conclusion canonical right-introduction (left-introduction) rule".
\end{convention}

\begin{exas}
\label{canonical rules examples}
We give some examples for canonical rules.
\bd
%


\item\noindent{\hskip-12 pt\bf Implication:}\
The two usual rules for implication are:
$$\tup{\{\Ra p_1\}, \{p_2\Ra\}} \ / \ p_1 \su p_2 \Ra
\mbox{\ \ \ and \ \ \ }
\{p_1 \Ra p_2 \} \ / \ \ \Ra p_1 \su p_2$$
Applications of these rules have the form:
\[
\derb{\g\Ra \psi}{\g,\fe \Ra E}
{\g,\psi \su \fe \Ra E}
\ \ \ \ \
\dera{\g,\psi \Ra \fe}
{\g \Ra \psi \su \fe}
\]
\item\noindent{\hskip-12 pt\bf Absurdity]:}\
In intuitionistic logic there is no
right-introduction rule for the absurdity constant $\perp$,
and there is exactly one left-introduction rule for it:
\begin{center}
$\tup{\emptyset,\emptyset} \ / \ \perp \ \Ra \  $
\end{center}
Applications of this rule provide new {\em axioms}:
\begin{center}
$\g,\perp \ \Ra E$
\end{center}
\item\noindent{\hskip-12 pt\bf Negation:}\
Unlike in \cite{AL10}, in this new framework it is possible to handle 
negation as a basic connective, using the following standard rules:
\begin{center}
$\tup{\{\Ra p_1 \}, \emptyset} \ / \ \neg p_1 \Ra \ $ \ and \
$\{p_1 \Ra \} \ / \ \Ra \neg p_1 \ $
\end{center}
Applications of these rules have the form:
\[
\dera{\g \Ra \psi} {\g,\neg \psi \Ra E}
\ \ \ \ \
\dera{\g,\psi \Ra } {\g \Ra \neg \psi}
\]
\item\noindent{\hskip-12 pt\bf Semi-Implication:}\
In \cite{gu09}  $\leadsto$ was introduced using
the following two rules:  
$$\tup{\{\Ra p_1 \}, \{p_2\Ra\}} \ / \ p_1 \leadsto p_2 \Ra
\mbox{\ \ \ and \ \ \ }
\{\Ra p_2 \} \ / \ \ \Ra p_1 \leadsto p_2$$
Applications of these rules have the form:
\[
\derb{\g\Ra \psi}{\g,\fe \Ra E}
{\g,\psi \leadsto \fe \Ra E}
\ \ \ \ \
\dera{\g \Ra \fe}
{\g \Ra \psi \leadsto \fe}
\]
\item\noindent{\hskip-12 pt\bf Affirmation:}\
Let the connective  $\affirm$ be defined using 
the following rules:
\begin{center}
$\tup{\emptyset, \{p_1 \Ra\}} \ / \ \affirm p_1  \Ra \ $ \ and \
$\{\Ra p_1\} \ / \ \Ra \affirm p_1 \ $
\end{center}
Applications of these rules have the form:
\[
\dera{\g, \fe \Ra E}
{\g,\affirm \fe \Ra E}
\ \ \ \ \
\dera{\g \Ra \fe}
{\g \Ra \affirm \fe}
\]
\item\noindent{\hskip-12 pt\bf Weak Affirmation:}\
Let the connective  $\waffirm$ be defined using
the following rules:

\begin{center}
$\tup{\{p_1 \Ra\},\emptyset} \ / \ \waffirm p_1 \Ra \ $ \ and \
$\{\Ra p_1\} \ / \ \Ra \waffirm p_1 \ $
\end{center}
Applications of these rules have the form:
\[
\dera{\g, \fe \Ra }
{\g,\waffirm \fe \Ra E}
\ \ \ \ \
\dera{\g \Ra \fe}
{\g \Ra \waffirm \fe}
\]
Note that the left-introduction rule for $\waffirm$ includes one hard negative 
premise, to which no right context can be added.
As a result, $\waffirm \fe \Ra \fe$ is not provable. 
\item\noindent{\hskip-12 pt\bf Bowen's connectives:}\
In \cite{Bowen}, Bowen introduced
 an extension of the basic intuitionistic calculus
with two new intuitionistic connectives\footnote
{He also presented ``neither-nor" connective, which we do not describe here, since 
this connective can be expressed by the four basic intuitionistic connectives.}.
He defined these connectives by the following canonical rules: 
\begin{center}
$\tup{\{p_2 \Ra p_1 \},\emptyset} \ / \ p_1 \not\subset p_2 \Ra$ \ and  \
$\{(p_1 \Ra),(\Ra p_2)\} \ / \ \Ra p_1 \not\subset p_2 \ $

\bs
$\tup{\{(\Ra p_1),(\Ra p_2)\}, \emptyset} / p_1 \mid p_2 \Ra$  \ and \
$\{p_1 \Ra \} / \Ra p_1 \mid p_2 \ $ \ \
$\{p_2 \Ra \} / \Ra p_1 \mid p_2 \ $
\end{center}
Applications of these rules have the form:
\[
\dera{\g, \psi \Ra \fe}
{\g,\fe \not\subset \psi\Ra E}
\ \ \ \ \
\derb{\g, \fe \Ra}{\g \Ra \psi}
{\g \Ra \fe \not\subset \psi}
\]
\[
\derb{\g \Ra \fe}{\g \Ra \psi}
{\g,\fe \mid \psi\Ra E}
\ \ \ \ \
\dera{\g, \fe \Ra}
{\g \Ra \fe \mid \psi}
\ \
\dera{\g, \psi \Ra}
{\g \Ra \fe \mid \psi}
\]


\ed
\end{exas}

\begin{defi}
\label{canonical system}
A non-strict single-conclusion sequent system 
is called {\em canonical} if it satisfies
the following conditions: 
\be
\item Its axioms are the sequents of the form $\fe\Ra\fe$.
\item Weakening and cut are among its rules.
Applications of these rules have the form:
\[
\dera{\g \Ra E}
{\g, \de \Ra E}
\ \
\dera{\g \Ra }
{\g \Ra \psi}
\ \ \ \ \
\derb{\g \Ra \fe}{\de,\fe \Ra E}
{\g,\de\Ra E}
\]
\item
Each of its other rules is 
either a canonical right-introduction rule 
or a canonical left-introduction rule.
\ee
\end{defi}

\begin{convention}
From now on, by {\em canonical system} 
we shall mean ``non-strict single-conclusion canonical system".
\end{convention}

\begin{defi}
\label{syntactic tcr}
Let ${\G}$ be a canonical system,
and let $\Ss \cup \{s\}$ be a set of sequents.
$\Ss\vd_{\G}^{seq} s$ iff there exists a derivation in ${\G}$ of $s$ from $\Ss$.
The sequents of $\Ss$ are called the {\em assumptions} (or {\em non-logical axioms}) of such a derivation.
\end{defi}


\begin{defi}
The etcr $\vd_{\G}$ 
which is induced by a canonical system ${\G}$ is defined by:
$\T\vd_{\G}E$ iff there exists a
finite $\g\suq\T$ such that $\vd_{\G}^{seq}\g\Ra E$.
\end{defi}

\begin{prop}
\label{structural finitary}
$\vd_{\G}$ is a structural and finitary etcr
for every canonical system $\G$.
\qed
\end{prop}

\begin{prop}
\label{reduction1}
$\T\vd_{\G} E$ iff
$\{\Ra\psi\st\psi\in\T\}\vd_{\G}^{seq}\Ra E$.
\qed
\end{prop}

We leave the easy proofs of the last two propositions to the reader.

\section{Consistency and Coherence}

Consistency (or non-triviality) is a crucial property of a deductive system. 
The goal of this section is to find a constructive criterion
for it in the framework of canonical systems.

\begin{defi}
\label{consistency}
A canonical system ${\G}$ is called {\em consistent} iff $\not\vd_{\G}^{seq} p_1 \Ra p_2$.
\end{defi}

\begin{prop}
\label{consistent proposition}
A canonical system $\G$ is consistent iff $\vd_{\G}$ is consistent.
\qed
\end{prop}

In multiple-conclusion canonical systems (\cite{AL05}), as well as in strict
single-conclusion  canonical systems (\cite{AL10}), consistency is equivalent to
coherence. Roughly speaking, a coherent system is a system in
which the rules cannot lead to new conflicts: the conclusions of
two rules can contradict each other only if their joint set of
premises is already inconsistent.  Next we
adapt this criterion to the present case:

\begin{defi}
\label{coherent-connective}
A set ${\mathcal{R}}$ of canonical rules for an $n$-ary connective $\dd$
is called {\em coherent} iff 
$S_1\cup S_2\cup S_3$ is classically inconsistent
whenever ${\mathcal{R}}$ contains 
both $\tup{S_1,S_2}/\dd(\ptil) \Ra \ $
and $S_3/\ \Ra \dd(\ptil)$. 
\end{defi}

\begin{rem}
It is known that a set of clauses is classically inconsistent
iff the empty clause can be derived from it using only cuts.
\end{rem}

\begin{exa}
Every connective introduced in Example \ref{canonical rules examples}, has a coherent set of rules.
For example, for the two rules
for implication we have $S_1=\{\ \Ra p_1\}$, $S_2=\{p_2\Ra\ \}$,
$S_3=\{p_1\Ra p_2\}$, and
$S_1\cup S_2\cup S_3$ is the classically inconsistent set
${\{(\ \Ra p_1) , (p_2\Ra\ ), (p_1\Ra p_2)\}}$ 
(from which the empty sequent can be derived using two cuts). 
For the two rules
for semi-implication we have $S_1=\{\ \Ra p_1\}$, $S_2=\{p_2\Ra\ \}$,
$S_3=\{\ \Ra p_2\}$, and
$S_1\cup S_2\cup S_3$ is the classically inconsistent set
${\{(\ \Ra p_1) , (p_2\Ra\ ), (\ \Ra p_2)\}}$ 
(from which the empty sequent can be derived using one cut). 
\end{exa}

\begin{defi}
\label{coherence}
A canonical system ${\G}$ 
is called {\em coherent} iff for each connective $\diamond$, the 
set of rules in $\G$ for $\diamond$ is coherent.
\end{defi}

Unfortunately, the next example shows that
in the present case coherence is {\em not}  necessary for consistency.

\begin{exa}
\label{G with circle}
Let ${\G}$ be a canonical system for a language which includes
 a single unary connective $\circ$, having
the following rules:
\begin{center}
$\tup{\emptyset,\{p_1 \Ra \}} \ / \ \circ p_1 \Ra \ $
\ and \
$\{p_1 \Ra \} \ / \  \ \Ra \circ p_1 $
\end{center}
Applications of these rules have the form:
\[
\dera{\g, \fe \Ra E}
{\g,\circ \fe \Ra E}
\ \ \ \ \
\dera{\g, \fe \Ra }
{\g \Ra \circ \fe}
\]
Obviously, ${\G}$ is not coherent.
However, 
it can easily be proved (using induction) that the only sequents
provable in $\G$ from no assumptions
are the sequents of the form $\g\Ra\psi$, where $\circ^n\psi\in\g$ for some $n\geq 0$
(here $\circ^0\psi=\psi$ and $\circ^{n+1}\psi=\circ\circ^{n}\psi$).
In particular, $p_1\Ra p_2$ is not provable in $\G$ from no assumptions, and so $\G$ is consistent.
\end{exa}

\ms
To overcome this difficulty, we define a stronger notion of consistency,
and show that in the context of non-strict canonical systems,
the coherence criterion is equivalent to this stronger notion.

\begin{defi}
\label{strong consistency}
A canonical system ${\G}$ is called {\em strongly consistent} 
iff ${(\Ra p_1),(p_2\Ra)\not\vd_{\G}^{seq}\Ra}$.
\end{defi}

\begin{prop}
\label{strong consistency -> consistency}
Every strongly consistent canonical system is also consistent.
\end{prop}
\proof
Let ${\G}$ be an inconsistent canonical system.
Then $\vd_{\G}^{seq} p_1 \Ra p_2$. 
Using the assumptions $(\Ra p_1),(p_2\Ra)$ and two cuts we get
$(\Ra p_1),(p_2\Ra)\vd_{\G}^{seq} \Ra$.
\qed

\medskip\noindent
The following derivation shows that the system from Example \ref{G with circle}
is not strongly consistent, and so strong consistency is indeed 
strictly stronger than consistency.

$$\infer[cut]{\Ra} {
	 { p_1\Ra} 
 	& \infer[cut]	{ \Ra p_1}   {
  		 \infer[\circ\Ra] {\circ p_1 \Ra p_1} {p_1 \Ra p_1}  & 
  		 \infer[\Ra\circ] { \Ra \circ p_1} {p_1 \Ra } } }$$

\noindent We note that strong consistency is a very natural demand
from a system: in strongly inconsistent systems it suffices to have
one provable sequent of the form $\psi\Ra$, and one provable sequent
of the form $\Ra\fe$, to make every sequent provable.

\begin{thm}
\label{strong consistency -> coherence}
Every strongly consistent canonical system is coherent.
\end{thm}
\proof
Let ${\G}$ be an incoherent canonical system.
This means that ${\G}$ includes two rules $\tup{S_1,S_2}/\dd(\ptil) \Ra \ $ and
$S_3/\ \Ra \dd(\ptil)$, such that the set of clauses
$S_1\cup S_2\cup S_3$ is classically satisfiable. Let $v$ be an assignment
in $\{t,f\}$ that satisfies all the clauses in $S_1\cup S_2\cup S_3$. 
Define a substitution $\sigma$ by:
\[\sigma(p)= \left\{
\begin{array}{ll}
p_1   & v(p)=t\\
p_2   & v(p)=f
\end{array}
\right.\] 
Since $v$ satisfies all the clauses in $S_1\cup S_2\cup S_3$,
for every ${\Pi\Ra E\in S_1\cup S_2\cup S_3}$ we have 
$p_2 \in \sigma(\Pi)$ or $p_1 \in \sigma(E)$.
Hence, every element of ${\sigma(S_1\cup S_2\cup S_3)}$ can be derived from ${(\Ra p_1),(p_2\Ra)}$ by weakening.
Now by applying the rules ${\tup{S_1,S_2}/\dd(\ptil) \Ra \ }$ and ${S_3/\ \Ra \dd(\ptil)}$ 
to these  sequents we get proofs from ${(\Ra p_1),(p_2\Ra)}$ of 
the sequents ${\Ra \sigma(\dd(\ptil))}$ and ${\sigma(\dd(\ptil))\Ra}$. 
That ${(\Ra p_1),(p_2\Ra)\vd_{\G}^{seq} \ \Ra \ }$ then follows using a cut.
\qed

\noindent
The last theorem implies that coherence is a necessary
demand from any acceptable canonical system ${\G}$. 
In the sequel (Corollary \ref{equivalences})
we show that coherence is also sufficient 
to ensure strong consistency.


\begin{rem}
Our coherence criterion can be proved to be equivalent (for fully-structural sequent systems) 
to the {\em reductivity} criterion defined in \cite{CT06}. 
However,  in the framework of \cite{CT06} a connective
essentially has infinitely many introduction rules,
while our framework makes it possible to convert
these infinite sets of rules into  finite ones. 
\end{rem}

\section{Semantics for Canonical Systems}
\label{semantic section}

In this section we generalize Kripke semantics for intuitionistic logic
to arbitrary coherent canonical systems. For this we
use {\em non-deterministic} Kripke frames and semiframes.

\begin{defi}
\label{persistent}
Let $\tup{W,\leq}$ be a nonempty partially ordered set.
Let $\U$ be a set of formulas.
A function $v:W \times \U\to \{t,f\}$ is called {\em persistent}
iff for every $a\in W$ and $\fe \in \U$,
 $v(a,\fe)=t$ implies that $v(b,\fe)=t$ for every $b\in W$
such that $a \leq b$.
\end{defi}

\begin{defi}
\label{semiframe}
Let $\U$ be a set of formulas closed under subformulas.
A {\em $\U$-semiframe} is a triple $\W=\tup{W, \leq,v}$
such that:
\be
\item $\tup{W,\leq}$ is a nonempty partially ordered set.
\item $v$ is a persistent function from $W \times {\U}$ to $\{t,f\}$.
\ee
When $\U=\F$ a $\U$-semiframe is also called an 
{\em $\lp$-frame}.
\end{defi}

\begin{rem}
\label{analytic remark}
To understand the need to consider semiframes, we note that 
to be useful and effective, a denotational semantics of 
a propositional logic should be {\em analytic}. 
This means that in order to determine
whether a sequent $s$ follows from a set  $\Ss$ of sequents,
it should be sufficient to consider {\em partial} valuations,
defined only on the set of subformulas of the formulas in
${\Ss}\cup\{s\}$. In the present case, such partial valuations are 
provided by semiframes.
\end{rem}

\begin{defi}
\label{sequent-satisfaction} 
Let ${\W=\tup{W, \leq,v}}$ be a $\U$-semiframe.
\be
\item A sequent ${\g\Ra E}$ is {\em locally true} in $a\in W$ 
iff $\g\cup E\suq\U$, and either ${v(a,\psi)=f}$ for some ${\psi\in\g}$, or 
$E=\{\fe\}$ and ${v(a,\fe)=t}$.
\item A sequent is {\em true} (or {\em absolutely true}) in $a\in W$ iff
it is locally true in every $b\geq a$.
\item ${\W}$ is a {\em model} of a sequent $s$ 
iff $s$ is true in every $a\in W$ (equivalently, if $s$ is locally true in every $a\in W$). 
It is a model of a set $\Ss$ of sequents if it is a model of every $s\in{\Ss}$.
\item ${\W}$ is a {\em model} of a formula $\fe$ iff $v(a,\fe)=t$ for every $a\in W$. 
It is a model of a theory $\T$ if it is a model of every $\fe\in\T$.
\ee
\end{defi}

\begin{rem}
From the point of view of local truth, a sequent 
is understood according to its classical interpretation as a disjunction
(either one of the formulas in its left side is ``false" or its right side is ``true").
On the other hand, the notion of absolute truth is based on viewing 
a sequent as expressing a real (constructive) entailment between its two sides.
Note that because of the persistence condition, 
for sequents of the form $\Ra\fe$ there is no difference between
local truth in $a$ or absolute truth in $a$.
Obviously, ${\W}$ is a model of such a sequent iff it is a model of $\fe$.
\end{rem}

Persistence is the only general condition which is satisfied by the semantics of
every coherent canonical system.
In addition, to every specific canonical system corresponds a set of constraints 
which are directly related to its set of canonical rules.
The idea is that a canonical rule for a connective $\dd$
imposes restrictions on the truth-values that can be 
assigned to $\diamond$-formulas.
Next we describe these restrictions.

\begin{defi}
\label{substitution satisfaction, fulfil and respect} 
Let $\W=\tup{W,\leq,v}$ be a $\U$-semiframe.
\be
\item 
An $\lp$-substitution $\sigma$ {\em (locally) satisfies} a sequent $\g\Ra E$ in 
$a\in W$ iff $\sigma(\g)\Ra\sigma(E)$ is (locally) true in $a$\footnote{When $E=\emptyset$, recall that $\sigma(\emptyset)=\emptyset$.}.
\item 
An $\lp$-substitution {\em fulfils} a canonical right-introduction rule in
$a \in W$ (with respect to $\W$) iff it satisfies in $a$ every premise of the rule.
\item
An $\lp$-substitution {\em fulfils} a canonical left-introduction rule in
$a \in W$ (with respect to $\W$) iff it satisfies in $a$ every hard premise of the rule,
and locally satisfies in $a$ every soft premise of the rule.
\item
Let $r$ be a canonical rule for an $n$-ary connective $\diamond$.
$\W$ {\em respects} $r$
iff for every $a\in W$ and every substitution $\sigma$:
if $\sigma$ fulfils $r$ in $a$ and $\sigma(\diamond(\ptil))\in\U$
then $\sigma$ locally satisfies conclusion of $r$ in $a$.
\ee
\end{defi}

\noindent Note that absolute truth is used for premises of right
introduction rules, as well as for hard premises of left introduction
rules.  Local truth is used only for soft premises of left
introduction rule.  This is the main difference between this semantics
and the one described in \cite{AL10} for the strict framework.  In
\cite{AL10}, the difference between absolute and local truth
corresponds to the syntactic distinction between definite and negative
sequents (absolute truth is used for definite premises, and local
truth is used for negative premises).  In the present case, since
negative sequents may also serve as premises of right introduction
rules and as hard premises of left introduction rules, this syntactic
distinction is irrelevant for the semantics definition.

\begin{rem}
Because of the persistence condition,
a definite sequent of the form
$\Ra \psi$ is satisfied in $a$ by $\sigma$ iff $v(a,\sigma(\psi))=t$.
\end{rem}

\begin{exas}
We describe the semantic effects of some rules
from Example \ref{canonical rules examples}.
\bd
\item\noindent{\hskip-12 pt\bf Negation:}\
An $\lp$-frame $\W=\tup{W, \leq,v}$
respects the rule $(\neg\Ra)$ if
${v(a,\neg\psi)=f}$ whenever $v(a,\psi)=t$. 
Because of the persistence condition, 
if ${v(b,\neg\psi)=f}$ for some $b\geq a$ then ${v(a,\neg\psi)=f}$.
And so, $\W$ respects $(\neg\Ra)$ if
${v(a,\neg\psi)=f}$ whenever $v(b,\psi)=t$ for some $b \geq a$.
It respects $(\Ra \neg)$ if $v(a,\neg\psi)=t$ whenever $v(b,\psi)=f$ 
for every $b \geq a$.
Hence the two rules together impose exactly the well-known
Kripke semantics for intuitionistic negation.


\item\noindent{\hskip-12 pt\bf Implication:}\
An $\lp$-frame $\W=\tup{W, \leq,v}$
respects the rule $(\su\Ra)$ iff for every $a\in W$,
$v(a,\fe\su\psi)=f$ whenever $v(b,\fe)=t$ for every $b \geq a$
and $v(a,\psi)=f$
(the latter -- because $\psi\Ra$ is an instance of a soft premise).
Because of the persistence condition, this is equivalent to
$v(a,\fe\su\psi)=f$ whenever $v(a,\fe)=t$ and $v(a,\psi)=f$.
Again by the persistence condition, $v(a,\fe\su\psi)=f$ iff 
$v(b,\fe\su\psi)=f$ for some $b\geq a$.
Hence, we get:
$v(a,\fe\su\psi)=f$ whenever there exists $b\geq a$
such that $v(b,\fe)=t$ and $v(b,\psi)=f$.
$\W$ respects $(\Ra\su)$ iff for every $a\in W$,
$v(a,\fe\su\psi)=t$ whenever for every $b\geq a$, either
$v(b,\fe)=f$ or $v(b,\psi)=t$.
Hence the two rules together impose exactly the well-known
Kripke semantics for intuitionistic implication (\cite{Kr65}).
It is easy to verify that the same applies to 
conjunction and disjunction, using the 
usual rules for these connectives.

\item\noindent{\hskip-12 pt\bf Semi-Implication:}\
An $\lp$-frame $\W=\tup{W, \leq,v}$
respects the rule $(\leadsto\Ra)$ under the same conditions
it respects $(\su\Ra)$. 
$\W$ respects $(\Ra\leadsto)$ iff for every $a\in W$,
${v(a,\fe\leadsto\psi)=t}$ whenever $v(a,\psi)=t$ (recall
that this is equivalent to $v(b,\psi)=t$ for every $b\geq a$).
Note that in this case the two rules for $\leadsto$ do not
always determine the value assigned to $\fe\leadsto\psi$:
if $v(a,\psi)=f$, and there is no $b\geq a$ such that $v(b,\fe)=t$ 
and $v(b,\psi)=f$,
then $v(a,\fe\leadsto\psi)$ is free to be either $t$ or $f$. So
the semantics of this connective is {\em non-deterministic}.



\item\noindent{\hskip-12 pt\bf Converse Non-Implication:}\
An $\lp$-frame ${\W=\tup{W, \leq,v}}$
respects the rule $(\not\subset\Ra)$ provided that
$v(a,\fe\not\subset\psi)=f$ whenever for every $b \geq a$
either $v(b,\fe)=t$ or $v(b,\psi)=f$. 
Because of the persistence condition, this is equivalent to
$v(a,\fe\not\subset\psi)=f$ if either there exists some 
$b \geq a$ such that $v(b,\fe)=t$, or if $v(b,\psi)=f$ for every $b \geq a$.
It respects $(\Ra \not\subset)$ if
$v(a,\fe\not\subset\psi)=t$ whenever
$v(b,\fe)=f$ and $v(b,\psi)=t$ for every $b \geq a$. 
Because of the persistence condition, this is equivalent to
$v(a,\fe\not\subset\psi)=t$ whenever $v(a,\psi)=t$ and
$v(b,\fe)=f$ for every $b \geq a$.
This implies that $v(a,\fe\not\subset\psi)$ is free
when $v(b,\fe)=f$ for every $b \geq a$, 
$v(a,\psi)=f$, and 
there exists $b \geq a$ such that $v(b,\psi)=t$.
For example, consider the following two 
$\{p_1,p_2,p_1\not\subset p_2\}$-semiframes:

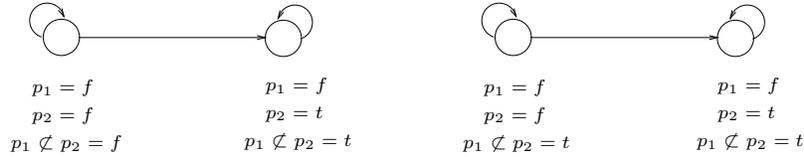
\bs

\begin{center}

\setlength{\unitlength}{0.00033333in}
\begingroup\makeatletter\ifx\SetFigFont\undefined%
\gdef\SetFigFont#1#2#3#4#5{%
  \reset@font\fontsize{#1}{#2pt}%
  \fontfamily{#3}\fontseries{#4}\fontshape{#5}%
  \selectfont}%
\fi\endgroup%
{\renewcommand{\dashlinestretch}{30}
\begin{picture}(6000,2389)(0,-10)
\put(802,1830){\ellipse{566}{566}}
\put(4267,1815){\ellipse{566}{566}}
\put(3990,1005){\makebox(0,0)[lb]{\smash{{\SetFigFont{7}{8.4}{\rmdefault}{\mddefault}{\updefault}$p_1=f$}}}}
\put(3990,570){\makebox(0,0)[lb]{\smash{{\SetFigFont{7}{8.4}{\rmdefault}{\mddefault}{\updefault}$p_2=t$}}}}
\put(3660,135){\makebox(0,0)[lb]{\smash{{\SetFigFont{7}{8.4}{\rmdefault}{\mddefault}{\updefault}$p_1\not\subset p_2=t$}}}}
\put(15,105){\makebox(0,0)[lb]{\smash{{\SetFigFont{7}{8.4}{\rmdefault}{\mddefault}{\updefault}$p_1\not\subset p_2=f$}}}}
\put(345,540){\makebox(0,0)[lb]{\smash{{\SetFigFont{7}{8.4}{\rmdefault}{\mddefault}{\updefault}$p_2=f$}}}}
\put(345,975){\makebox(0,0)[lb]{\smash{{\SetFigFont{7}{8.4}{\rmdefault}{\mddefault}{\updefault}$p_1=f$}}}}
\put(4449.500,2063.559){\arc{541.556}{3.3895}{7.5171}}
\path(4215.385,2250.392)(4187.000,2130.000)(4268.700,2222.871)
\path(3863.990,1799.041)(3984.000,1829.000)(3864.010,1859.041)
\path(3984,1829)(1085,1830)
\put(579.500,2093.559){\arc{541.556}{1.9077}{6.0353}}
\path(760.300,2252.871)(842.000,2160.000)(813.615,2280.392)
\end{picture}
}
\ \ \ \ \ 
\begin{picture}(6000,2389)(0,-10)
\put(802,1830){\ellipse{566}{566}}
\put(4267,1815){\ellipse{566}{566}}
\put(3990,1005){\makebox(0,0)[lb]{\smash{{\SetFigFont{7}{8.4}{\rmdefault}{\mddefault}{\updefault}$p_1=f$}}}}
\put(3990,570){\makebox(0,0)[lb]{\smash{{\SetFigFont{7}{8.4}{\rmdefault}{\mddefault}{\updefault}$p_2=t$}}}}
\put(3660,135){\makebox(0,0)[lb]{\smash{{\SetFigFont{7}{8.4}{\rmdefault}{\mddefault}{\updefault}$p_1\not\subset p_2=t$}}}}
\put(15,105){\makebox(0,0)[lb]{\smash{{\SetFigFont{7}{8.4}{\rmdefault}{\mddefault}{\updefault}$p_1\not\subset p_2=t$}}}}
\put(345,540){\makebox(0,0)[lb]{\smash{{\SetFigFont{7}{8.4}{\rmdefault}{\mddefault}{\updefault}$p_2=f$}}}}
\put(345,975){\makebox(0,0)[lb]{\smash{{\SetFigFont{7}{8.4}{\rmdefault}{\mddefault}{\updefault}$p_1=f$}}}}
\put(4449.500,2063.559){\arc{541.556}{3.3895}{7.5171}}
\path(4215.385,2250.392)(4187.000,2130.000)(4268.700,2222.871)
\path(3863.990,1799.041)(3984.000,1829.000)(3864.010,1859.041)
\path(3984,1829)(1085,1830)
\put(579.500,2093.559){\arc{541.556}{1.9077}{6.0353}}
\path(760.300,2252.871)(842.000,2160.000)(813.615,2280.392)
\end{picture}

\end{center}

\ms\noindent
While there is no difference between these two semi-frames 
with respect to atomic formulas, 
the truth-values assigned to $p_1\not\subset p_2$ 
in one of their two worlds are different.
Now both semiframes respect the two rules of $\not\subset$.
Hence the semantics of this connective is 
non-deterministic.\footnote{Note that no semantic characterizations 
for ``converse non-implication" and ``not both"
were presented in \cite{Bowen}, where these connectives were first introduced.}

\item\noindent{\hskip-12 pt\bf Not Both:}\
An $\lp$-frame $\W=\tup{W, \leq,v}$
respects the rule $(\mid\Ra)$ if
${v(a,\fe\mid\psi)=f}$ whenever 
$v(b,\fe)=t$ and $v(b,\psi)=t$ for every $b \geq a$. 
Because of the persistence condition, this is equivalent to
$v(a,\fe\mid\psi)=f$ whenever
$v(b,\psi)=v(b,\fe)=t$
for some $b \geq a$.
It respects $(\Ra \mid)_1$ if
$v(a,\fe\mid\psi)=t$ whenever
$v(b,\fe)=f$  for every $b \geq a$.
It respects $(\Ra \mid)_2$ if
$v(a,\fe\mid\psi)=t$ whenever 
$v(b,\psi)=f$ for every $b \geq a$.
This implies that $v(a,\fe\mid\psi)$ is free
when there exist $b_1,b_2\geq a$ such that
$v(b_1,\fe)=v(b_2,\psi)=t$,
but there does not exist $b\geq a$ such that
$v(b,\fe)=v(b,\psi)=t$
(this is possible because the order relation does not have to be linear).
Again, the induced semantics is non-deterministic.
\item\noindent{\hskip-12 pt\bf Affirmation:}\
An $\lp$-frame $\W=\tup{W, \leq,v}$
respects the rule $(\affirm\Ra)$ if
${v(a,\affirm\psi)=f}$ whenever ${v(a,\psi)=f}$. 
It respects ${(\Ra \affirm)}$ if
$v(a,\affirm\psi)=t$ whenever ${v(a,\psi)=t}$.
This means that for every ${a \in W}$,
$v(a,\affirm\psi)$ simply equals ${v(a,\psi)}$.
\item\noindent{\hskip-12 pt\bf Weak Affirmation:}\
An $\lp$-frame $\W=\tup{W, \leq,v}$
respects the rule $(\waffirm\Ra)$ if
${v(a,\waffirm\psi)=f}$ whenever $v(b,\psi)=f$
for every $b \geq a$.
It respects $(\Ra \waffirm)$ if
$v(a,\waffirm\psi)=t$ whenever $v(b,\psi)=t$
for every $b\geq a$.
Because of the persistence condition, this is equivalent to
$v(a,\waffirm\psi)=t$ whenever $v(a,\psi)=t$.
This implies that $v(a,\waffirm\psi)$ is free
when $v(a,\psi)=f$ and $v(b,\psi)=t$ 
for some $b \geq a$.
Again, we obtain non-deterministic semantics.
\ed
\end{exas}


\begin{defi}
\label{G-legal}
Let ${\G}$ be a canonical system.
A $\U$-semiframe is {\em ${\G}$-legal} iff it respects all the canonical rules of ${\G}$.
\end{defi}

\indent
We can now give the definition of the semantic relations induced by a canonical system:

\begin{defi}
Let ${\G}$ be a coherent canonical system,
and let $\Ss \cup \{s\}$ be a set of sequents.
${\Ss\vD_\G^{seq} s}$ iff every ${\G}$-legal $\lp$-frame which is
a model of $\Ss$ is also a model of ${s}$.
\end{defi}

\begin{defi}
Let ${\G}$ be a coherent canonical system. 
The semantic etcr $\vD_\G$ between {\em formulas}
which is induced by ${\G}$ is defined by:
$\T\vD_\G E$ iff every ${\G}$-legal $\lp$-frame which is 
a model of $\T$ is also a model of $E$.
\end{defi}

\noindent
Again we have:

\begin{prop}
\label{reduction2}
$\T\vD_{\G} E$ iff $\{\Ra\psi\st\psi\in\T\}\vD_{\G}^{seq}\Ra E$.
\qed
\end{prop}

\section{Soundness, Completeness, Cut-elimination}
In this section we show that the syntactic and semantic consequence relations between sequents
which are induced by a given coherent canonical system are identical.
In addition, we present a semantic proof of cut-elimination for arbitrary coherent canonical systems.
There are a lot of similarities between the proofs of this section and the corresponding proofs in 
\cite{AL10}. However, the proofs in \cite{AL10} correspond to different definitions,
and so, for the sake of completeness, we include here the full proofs.

\begin{thm}
\label{soundness}
Every coherent canonical system ${\G}$
is strongly sound with respect to the semantics
of ${\G}$-legal frames. In other words:
If $\Ss\vd_{\G}^{seq}s$ then $\Ss\vD_{\G}^{seq}s$.
\end{thm}

\sloppy
\proof
Assume that ${\Ss\vd_\G^{seq}s}$, 
and ${\W=\tup{W, \leq,v}}$ is a ${\G}$-legal model of $\Ss$. 
We show that $s$ is locally true in every  ${a\in W}$.
Since the axioms of ${\G}$ and the assumptions of $\Ss$ trivially
have this property, and the cut and weakening rules obviously preserve it,
it suffices to show that the property of being locally true in every $a\in W$
is also preserved by applications of the logical rules of ${\G}$.
\bi
\item 
Suppose ${\g\Ra \sigma(\dd(\ptil))}$
is derived from ${\{\g,\sigma(\Pi_i) \Ra \sigma(E_i)\}_{1 \leq i \leq m}}$
using the 
rule
${r=\{\Pi_i \Ra E_i\}_{1 \leq i \leq m} / \Ra \dd(\ptil)}$.
Assume that all the premises of this application have the required property.
We show that so does its conclusion. 
Let ${a\in W}$. 
If ${v(a,\psi)=f}$ for some ${\psi\in\g}$, then obviously  
${\g\Ra \sigma(\dd(\ptil))}$ is locally true in ${a}$. 
Assume otherwise. 
Then the persistence condition implies that
${v(b,\psi)=t}$ for every ${\psi\in\g}$ and ${b\geq a}$.
Thus our assumption concerning the sequents
${\{\g,\sigma(\Pi_i) \Ra \sigma(E_i)\}_{1 \leq i \leq m}}$ entails
that for every ${b\geq a}$ and ${1 \leq i \leq m}$, either ${v(b,\psi)=f}$ for
some ${\psi\in\sigma(\Pi_i)}$, or $E_i=\{q_i\}$ 
(i.e. $E_i$ is not empty) and ${v(b,\sigma(q_i))=t}$. 
It follows that for ${1 \leq i \leq m}$, ${\Pi_i \Ra E_i}$
is satisfied in ${a}$ by $\sigma$.
Thus, $\sigma$ fulfils $r$ in $a$.
Since ${\W}$ respects $r$, 
it follows that ${v(a,\sigma(\dd(\ptil)))=t}$.
\item
Now we deal with left-introduction rules.
Suppose ${\g, \sigma(\dd(\ptil))\Ra E}$ is derived from
${\{\g,\sigma(\Pi_i) \Ra \sigma(E_i)\}_{1 \leq i \leq m}}$ 
and ${\{\g,\sigma(\Sigma_i) \Ra E\}_{1 \leq i \leq k}}$,
using the left-introduction rule 
${r=\tup{\{\Pi_i \Ra E_i\}_{1 \leq i \leq m}, \{\Sigma_i \Ra \}_{1 \leq i \leq k}}/\dd(\ptil)\Ra}$.
Assume that all the premises of this application have the required property.
We show that so does its conclusion. 
Let ${a\in W}$. 
If ${v(a,\psi)=f}$ for some ${\psi\in\g}$ 
or ${E=\{\theta\}}$ and ${v(a,\theta)=t}$, then we are done. 
Assume otherwise. 
Then $E$ is either empty or ${E=\{\theta\}}$ and ${v(a,\theta)=f}$,
and (by the persistence condition)
${v(b,\psi)=t}$ for every ${\psi\in\g}$ and ${b\geq a}$.
Thus our assumption concerning the sequents
${\{\g,\sigma(\Pi_i) \Ra \sigma(E_i)\}_{1 \leq i \leq m}}$ entails
that for every ${b\geq a}$ and ${1 \leq i \leq m}$, either ${v(b,\psi)=f}$ for
some ${\psi\in\sigma(\Pi_i)}$, or ${E_i=\{q_i\}}$ and ${v(b,\sigma(q_i))=t}$.
This immediately implies that the hard premises of $r$ are satisfied in ${a}$ by $\sigma$. 
Since $E$ is either empty or ${E=\{\theta\}}$
and ${v(a,\theta)=f}$, our assumption concerning
${\{\g,\sigma(\Sigma_i) \Ra E\}_{1 \leq i \leq k}}$ entails
that for every ${1 \leq i \leq k}$,
${v(a,\psi)=f}$ for some ${\psi\in\sigma(\Sigma_i)}$.
Hence the soft premises of $r$ are locally satisfied in ${a}$ by $\sigma$.
Thus, $\sigma$ fulfils $r$ in $a$.
Since ${\W}$ respects $r$, 
it follows that ${v(a,\sigma(\dd(\ptil)))=f}$.
\qed\ei
\fussy

\ms
For the converse, we define {\em $\Ss$-proofs} and prove the following key result.

\begin{defi}
\label{S-proof}
Let ${\Ss}$ be sets of sequents.
A proof $P$ in a canonical system is called an {\em $\Ss$-proof} iff
the cut formula of every cut in $P$ occurs in $\Ss$.
\end{defi}

\begin{thm}
\label{key}
Let ${\G}$ be a coherent canonical system in $\lp$,
and let $\Ss\cup \{s\}$ be a set of sequents in $\lp$. 
Then either there is an $\Ss$-proof of $s$ from $\Ss$ in $\G$,
or there is a ${\G}$-legal $\lp$-frame 
which is model of $\Ss$, but not a model of $s$.
\end{thm}
\proof
Assume that ${s=\g_0\Ra E_0}$ does not have an
$\Ss$-proof from $\Ss$ in ${\G}$.
We construct a ${\G}$-legal $\lp$-frame ${\W}$ which
is a model of $\Ss$ but not of ${s}$.
Let ${\U}$ be the set of subformulas of ${\Ss\cup \{s\}}$.
Given a subset $E$ of $\U$ which is either a singleton or empty, 
call a theory ${\T\suq\U}$ {\em $E$-maximal}
if there is no finite ${\g\suq\T}$ such that
${\g\Ra E}$ has an $\Ss$-proof from $\Ss$,
but every proper extension ${\T^\prime\suq\U}$
of $\T$ contains such  a finite subset $\g$.
Obviously, if ${\g\cup  E \suq\U}$
and ${\g\Ra E}$ has no $\Ss$-proof from $\Ss$,
then ${\g}$ can be extended to a theory ${\T\suq\U}$ which is $E$-maximal.
In particular: ${\g_0}$ can be extended to a ${E_0}$-maximal theory ${\T_0}$.

\ms

Now let ${\W=\tup{W,\suq,v}}$, where:
\bi
\item ${W}$ is the set of all extensions of ${\T_0}$ 
in ${\U}$ which are $E$-maximal for some ${E\suq\U}$
(recall that $E$ is either singleton or empty).
\item ${v}$ is defined inductively as follows. 
For atomic formulas:
 \[v(\T, p)= \left\{
\begin{array}{ll}
t & p\in\T\\
f & p\not\in\T
\end{array}
\right.\]
Suppose ${v(\T, \psi_i)}$ has been defined for every ${\T\in W}$
and ${1 \leq i \leq n}$.
\\We let ${v(\T, \dd(\psitil))=t}$ iff
at least one of the following holds 
with respect to the semiframe constructed so far:
\be
\item There exists a right-introduction rule for $\dd$ 
which is fulfilled in $\T$ by a substitution $\sigma$ such that
${\sigma(p_i)=\psi_i}$ (${1\leq i\leq n}$).
\item

\sloppy

 ${\dd(\psitil) \in \T}$, and
there do not exist ${\T^\prime\in W}$ 
and a left-introduction rule $r$ for $\dd$, such that ${\T\suq\T^\prime}$,
and $r$ is fulfilled in ${\T^\prime}$ by a substitution $\sigma$ such that
${\sigma(p_i)=\psi_i}$ (${1\leq i\leq n}$).

\fussy
\ee
\ei

\smallskip
\noindent
First we prove that ${\W}$ is an $\lp$-frame:
\bi

\item ${W}$ is not empty because ${\T_0\in W}$.

\item We prove by structural induction that ${v}$ is persistent:\\
For atomic formulas ${v}$ is trivially persistent since the
order is ${\suq}$.\\
Assume that ${v}$ is persistent for ${\psitil}$.
We prove its persistence for ${\dd(\psitil)}$.
So assume  that ${v(\T,\dd(\psitil))=t}$ and ${\T\suq\T^*}$.
By the definition of $v$ there are two possibilities:
\be
\item There exists a right-introduction rule for $\dd$ 
which is fulfilled in $\T$ by a substitution $\sigma$ such that
${\sigma(p_i)=\psi_i}$ (${1\leq i\leq n}$).
This is also trivially true in ${\T^*}$, and so
${v(\T^*,\dd(\psitil))=t}$.
\item 

\sloppy
${\dd(\psitil) \in \T}$, and
there do not exist ${\T^\prime\in W}$ 
and a left-introduction rule $r$  for $\dd$, such that ${\T\suq\T^\prime}$, 
 and $r$  is fulfilled in ${\T^\prime}$ by a substitution $\sigma$ such that
${\sigma(p_i)=\psi_i}$ (${1\leq i\leq n}$).
Then ${\dd(\psitil) \in \T^*}$ (since ${\T\suq\T^*}$),
and there cannot exist ${\T^\prime\in W}$ 
and a left-introduction rule $r$ for $\dd$, such that ${\T^*\suq\T^\prime}$,
and $r$ is fulfilled in ${\T^\prime}$
by such a substitution $\sigma$ (otherwise the same would hold for $\T$).
Hence ${v(\T^*,\dd(\psitil))=t}$ in this case too.

\fussy

\ee
\ei

\smallskip
\noindent
Next we prove that ${\W}$ is ${\G}$-legal:
\be
\item The right-introduction rules are directly respected by the first
  condition in the definition of $v$.
\item Let $r$ be a left-introduction rule for $\dd$, and let $\T\in W$.
Suppose that $r$ is fulfilled in $\T$ by a substitution $\sigma$, 
such that ${\sigma(p_i)=\psi_i}$ (${1\leq i\leq n}$).
Then neither of the conditions
under which $v(\T,\dd(\psitil))=t$ can hold: 
\begin{enumerate}[(a)]
\item The second condition explicitly excludes the option that 
$r$ is fulfilled by $\sigma$
(in any ${\T^\prime\in W}$ such that
 ${\T\suq\T^\prime}$, including $\T$ itself).
\item The first condition cannot be met because the coherence of $\G$ does 
not allow the sets of premises (of a right-introduction rule and a left-introduction rule for the same connective) 
to be locally satisfied together. 
Hence the two rules cannot be both fulfilled
by the same substitution in the same element of $W$.
To see this, assume by way of contradiction that ${S_1}$ and ${S_2}$ are
the sets of premises of a left-introduction rule for $\dd$, ${S_3}$
is the set of premises of a right-introduction rule for $\dd$, 
and there exists ${\T\in W}$
in which the three sets of premises are locally satisfied by a substitution 
$\sigma$ such that ${\sigma(p_i)=\psi_i}$ (${1\leq i\leq n}$). 
Let ${u}$ be an assignment in ${\{t,f\}}$ 
in which ${u(p_i)=v(\T,\psi_i)}$.
Since $\sigma$ locally satisfies in $\T$ the three sets of premises, 
${u}$ classically satisfies ${S_1}$, ${S_2}$ and ${S_3}$. 
This contradicts the coherence of ${\G}$.
\ee
It follows that $v(\T,\dd(\psitil))=f$, as required.
\ee

\smallskip\noindent
It remains to prove that
${\W}$ is a model of $\Ss$ but not of ${s}$.
For this we first prove that the following hold
for every ${\T\in W}$ and every formula ${\psi\in\U}$:

\begin{enumerate}[\bf(a):]
\item If  ${\psi \in \T}$ then ${v(\T,\psi)=t}$.
\item If $\T$ is ${\{\psi\}}$-maximal then ${v(\T,\psi)=f}$.
\ee

\noindent
We prove {\bf (a)} and {\bf (b)} together
by a simultaneous induction on the complexity of $\psi$.
For atomic formulas they easily follow from the definition of $v$,
and the fact that ${p\Ra p}$ is an axiom.
For the induction step, 
assume that {\bf (a)} and {\bf (b)}  hold for ${\psitil\in\U}$. 
We prove them for ${\dd(\psitil)\in\U}$. 

\bi
\item Assume that ${\dd(\psitil)\in\T}$, but ${v(\T,\dd(\psitil))=f}$.
  By the definition of $v$, since ${\dd(\psitil)\in\T}$ there should
  exist ${\T^\prime\in W}$, ${\T\suq\T^\prime}$, and a
  left-introduction rule, ${r=\tup{\{\Pi_i \Ra E_i\}_{1 \leq i \leq
        m},\{\Sigma_i \Ra \}_{1 \leq i \leq k}} / \dd(\ptil) \Ra}$,
  fulfilled in ${\T^\prime}$ by a substitution $\sigma$ such that
  ${\sigma(p_i)=\psi_i}$ (${1\leq i\leq n}$).  As $\sigma$ locally
  satisfies in ${\T^\prime\!}$ every sequent in ${\{\Sigma_i\Ra\}_{1
      \leq i \leq k}}$, then for every ${1 \leq i \leq k}$ there
  exists ${\psi_{j_i}\in\sigma(\Sigma_i)}$ with
  ${v(\T^\prime,\psi_{j_i})=f}$.  By the induction hypothesis this
  implies that for every ${1 \leq i\leq k}$, there exists
  ${\psi_{j_i}\in\sigma(\Sigma_i)}$ such that
  ${\psi_{j_i}\notin\T^\prime}$.  Let $E$ be the set for which
  ${\T^\prime}$ is maximal.  Then for every ${1 \leq i \leq k}$ there
  is a finite ${\de_i\suq\T^\prime}$ such that ${\de_i,\psi_{j_i}\Ra E}$
  has an $\Ss$-proof from $\Ss$, and therefore
  ${\de_i,\sigma(\Sigma_i)\Ra E}$ has such a proof.  This in turn
  implies that there must exist ${1 \leq i_0 \leq m}$ such that
  ${\g,{\sigma(\Pi_{i_0})\Ra \sigma(E_{i_0})}}$ has no $\Ss$-proof
  from $\Ss$ for any finite ${\g\suq\T^\prime}$.  Indeed, if such a
  proof exists for every ${1 \leq i \leq m}$, we would use the $k$
  proofs of ${\de_i,\sigma(\Sigma_i)\Ra E}$ for ${1 \leq i \leq k}$,
  the $m$ proofs for ${\g_i,\sigma(\Pi_i)\Ra \sigma(E_i)}$ for ${1
    \leq i \leq m}$, some trivial weakenings, and the
  left-introduction rule $r$ to get an $\Ss$-proof from $\Ss$ of the
  sequent ${\cup_{i=1}^{i=k}\de_i,\cup_{i=1}^{i=m}\g_i,\dd(\psitil)\Ra
    E}$.  Since ${\dd(\psitil)\in\T}$, this would contradict the
  $E$-maximality of ${\T^\prime}$.  Using this ${i_0}$, extend
  ${\T^\prime\cup\sigma(\Pi_{i_0})}$ to a ${\sigma(E_{i_0})}$-maximal
  theory ${\T^{\prime\prime}}$.  By the induction hypothesis,
  ${v(\T^{\prime\prime},\psi)=t}$ for every
  ${\psi\in\sigma(\Pi_{i_0})}$, and if ${E_{i_0}=\{q\}}$
  (i.e. $E_{i_0}$ is not empty) then
  ${v(\T^{\prime\prime},\sigma(q))=f}$.  Since
  ${T^\prime\suq\T^{\prime\prime}}$, this contradicts the fact that
  $\sigma$ satisfies ${\Pi_{i_0}\Ra E_{i_0}}$ in ${\T^\prime}$.

\item Assume that $\T$ is ${\{\dd(\psitil)\}}$-maximal,
but that ${v(\T,\dd(\psitil))=t}$. 
Obviously, ${\dd(\psitil)\notin\T}$ 
(because ${\dd(\psitil)\Ra\dd(\psitil)}$ is an axiom). 
Hence 
there exists a right-introduction rule, 
${r=\{\Pi_i \Ra E_i\}_{1 \leq i \leq m} / \Ra \dd(\ptil)}$, 
which is fulfilled in ${\T}$
by a substitution $\sigma$ such that
${\sigma(p_i)=\psi_i}$ (${1\leq i\leq n}$).
As in the previous case, there must exist ${1 \leq i_0 \leq m}$
such that ${\g,\sigma(\Pi_{i_0})\Ra \sigma(E_{i_0})}$ has no
$\Ss$-proof from $\Ss$ for any finite ${\g\suq\T}$ (if
such a proof exists for every ${1 \leq i \leq m}$ with finite
${\g_i\suq\T}$ than we could have an $\Ss$-proof from $\Ss$ of 
${\cup_{i=1}^{i=m}\g_i\Ra\dd(\psitil)}$ 
using the ${m}$ proofs of ${\g_i,\sigma(\Pi_i)\Ra \sigma(E_i)}$, 
some weakenings and $r$). 
Using this ${i_0}$, extend ${\T\cup\sigma(\Pi_{i_0})}$
to a ${\sigma(E_{i_0})}$-maximal theory ${\T^\prime}$.
By the induction hypothesis
${v(\T^\prime,\psi)=t}$ for every ${\psi\in\sigma(\Pi_{i_0})}$, and
if ${E_{i_0}=\{q\}}$ (i.e. $E_{i_0}$ is not empty) then ${v(\T^\prime,\sigma(q))=f}$. 
Since ${\T\suq\T^\prime}$, this contradicts
the fact that $\sigma$ satisfies ${\Pi_{i_0}\Ra E_{i_0}}$ in $\T$.
\ei

\ms

\noindent
Next we note that {\bf (b)} can be strengthened as follows:

\bd
\item[(c)] If ${\psi\in\U}$, ${\T\in W}$ and there is no finite ${\g\suq\T}$ such that
${\g\Ra\psi}$ has an $\Ss$-proof from $\Ss$, then ${v(\T,\psi)=f}$.
\ed
\noindent Indeed, under these conditions $\T$ can be extended
to a $\{\psi\}$-maximal theory $\T^\prime$. Now $\T^\prime\in W$,
$\T\suq\T^\prime$, and by {\bf (b)}, $v(\T^\prime,\psi)=f$. 
Hence also ${v(\T,\psi)=f}$.

Now {\bf (a)} and {\bf (b)} together imply  that
${v(\T_0,\psi)=t}$ for every ${\psi\in\g_0\suq\T_0}$, and
if $E_0={\{\theta\}}$ (i.e. $E_0$ is not empty) then ${v(\T_0,\theta)=f}$. 
Hence ${\W}$ is not a model of ${s}$.
We end the proof by showing that ${\W}$ is a model of $\Ss$. 
So let ${\psitil\Ra E\in \Ss}$ and let ${\T\in W}$,
where $\T$ is $F$-maximal.
Assume by way of contradiction that
${\psitil\Ra E}$ is not locally true in $\T$.
Therefore, ${v(\T,\psi_i)=t}$ for $1\leq i\leq n$. 
By {\bf (c)}, for every $1\leq i\leq n$ 
there is a finite ${\g_i\suq\T}$ such that ${\g_i\Ra\psi_i}$
has an $\Ss$-proof from $\Ss$. 
Now, there are two cases:
\be
\item Assume ${E=\{\theta\}}$. 
Since ${\psitil\Ra \theta}$ is not locally true in $\T$, ${v(\T,\theta)=f}$.
This implies (by {\bf (a)}) that ${\theta \notin \T}$. 
Since $\T$ is $F$-maximal,
it follows that there is a finite ${\de\suq\T}$ such that
${\de,\theta\Ra F}$ has an $\Ss$-proof from $\Ss$.
Now from ${\g_i\Ra\psi_i}$ ($1\leq i\leq n$), ${\de,\theta\Ra F}$,
and ${\psitil\Ra\theta}$ one can infer
$\g_1\til\g_n,\de\Ra F$ by $n+1$ $\Ss$-cuts 
(on $\psitil$ and $\theta$). 
Hence, 
$\g_1\til\g_n,\de\Ra F$ has an $\Ss$-proof from $\Ss$.
\item Assume $E$ is empty.
$\g_1\til\g_n\Ra $ follows from 
the sequents ${\g_i\Ra\psi_i}$ ($1\leq i\leq n$) and ${\psitil\Ra }$ 
by $n$ $\Ss$-cuts (on $\psitil$). 
Using weakening (if $F$ is not empty), it follows that 
$\g_1\til\g_n\Ra F$ has an $\Ss$-proof from $\Ss$.
\ee 
In both cases we showed an $\Ss$-proof from $\Ss$ of a sequent of 
the form $\g \Ra F$, where $\g\suq\T$. 
This contradicts the $F$-maximality of $\T$.
\qed

\begin{rem}
This proof suggests that weakening on the right side of sequents can be limited
to apply only to negative sequents of the set of assumptions of the derivation.
Recall that by proposition \ref{reduction1}, 
$\T\vd_{\G} E$ iff $\{\Ra\psi\st\psi\in\T\}\vd_{\G}^{seq}\Ra E$.
Thus if one is only interested in consequence relations
 between formulas, there are no negative 
sequents in the set of assumptions, and so the right weakening rule is superfluous.
\end{rem}

\sloppy
\begin{thm}[Soundness and Completeness]
\label{completeness}
Every coherent canonical system ${\G}$
is strongly sound and complete with respect to the semantics
of ${\G}$-legal frames. In other words:
\be
\item $\Ss\vd_{\G}^{seq}s$ iff $\Ss\vD_{\G}^{seq}s$.
\item $\T\vd_{\G} E$ iff $\T\vD_{\G} E$.
\ee
\end{thm}

\fussy
\proof
($1$) is immediate from Theorem \ref{key} and Theorem \ref{soundness}.
($2$) follows from ($1$) using the reductions given 
in Proposition \ref{reduction1} and Proposition \ref{reduction2}.
\qed

\begin{cor}[Compactness]
Let ${\G}$ be a coherent canonical system.
If $\Ss\vD^{seq}_{{\G}}s$ 
then there exists a finite $\Ss^{\prime}\suq\Ss$ such that 
$\Ss^{\prime}\vD^{seq}_{{\G}}s$.
\qed
\end{cor}

We use Theorem \ref{key} to prove a general cut-elimination theorem.

\begin{defi}
\label{cut elimination}
Let $s$ be a sequent, ${\Ss}$ be a set of sequents,
and ${\G}$ be a canonical system.
\be
\item ${\G}$ admits cut-elimination iff whenever ${\vd_\G^{seq}s}$, there exists
a proof of ${s}$ without cuts (i.e. there exists a $\emptyset$-proof).
\item (\cite{Av93}) ${\G}$ admits strong
cut-elimination iff whenever  ${\Ss\vd_\G^{seq}s}$, there exists
an $\Ss$-proof of ${s}$ from $\Ss$.
\ee
\end{defi}

Notice that cut-elimination is a special case of strong cut-elimination with an empty $\Ss$.
Also notice that by cut-elimination we mean here just the existence of proofs without (certain forms
of) cuts, rather than an algorithm to transform a given proof to a cut-free one (for
the assumption-free case the term {\em cut-admissibility} is sometimes used).

\begin{thm}[General Strong Cut-Elimination Theorem]
\label{cut-elimination}
Every coherent canonical system ${\G}$
admits strong cut-elimination.
\end{thm}
\proof
Assume ${\Ss\vd_\G^{seq}s}$.
By  Theorem \ref{completeness}, ${\Ss\vD_\G^{seq}s}$,
and so there does not exist a ${\G}$-legal $\lp$-frame 
which is model of $\Ss$, but not a model of $s$.
By Theorem \ref{key}, there is an $\Ss$-proof of $s$ from $\Ss$.
\qed

\begin{rem}
\label{hyper-cut}
In \cite{Av93}, a strengthening of the cut-elimination theorem was suggested
for Gentzen's original systems for classical logic. 
The notion of a {\em hyper-resolution} rule (or {\em hyper-cut} rule) was defined,
and it was proven that this special kind of cuts is the only one needed in derivations
of a sequent from a non-empty set of sequents.
Following the proof of Theorem \ref{key}, we can show the same in the present case.
Let {\em hyper-cut}$_1$ and {\em hyper-cut}$_2$ be the rules which allow the following two derivations:
\[
\derc{\psitil\Ra \theta}{\g_1\Ra\psi_1 \ \ \ldots \ \ \g_n\Ra\psi_n}{\de,\theta\Ra F}
{\g_1 \til \g_n,\de\Ra F}
\]
\[
\derb{\psitil\Ra}{\g_1\Ra\psi_1 \ \ \ldots \ \ \g_n\Ra\psi_n}
{\g_1 \til \g_n\Ra}
\]
Call $\psitil \Ra E$, where $E=\{\theta\}$ in the 
first derivation and empty in the second,
the {\em nucleus} of the rule.
The last theorem can be strengthened as follows:
if ${\Ss\vd_\G^{seq}s}$,
then there exists a proof of $s$ from $\Ss$,
which uses only axioms, canonical rules,
weakenings and hyper-cuts with elements of $\Ss$ as nuclei.
\end{rem}

\begin{cor}
\label{equivalences}
The conditions below are equivalent for a 
canonical system ${\G}$:
\be
\item ${\G}$ is strongly consistent.
\item ${\G}$ is coherent.
\item ${\G}$ admits strong cut-elimination.
\ee
\end{cor}
\proof
(1) implies (2) by Theorem \ref{strong consistency -> coherence}. 
(2) implies (3) by Theorem \ref{cut-elimination}. 
Finally, in a canonical system
the only sequents which are provable from 
$\{(\Ra p_1),(p_2\Ra)\}$ using only cuts on $p_1$ or $p_2$ are: 
axioms, sequents of the form $\g\Ra p_1$,
sequents of the form $\g,p_2\Ra E$, and sequents that contain a non-atomic formula.
Thus there is no way to derive $\Ra$ from $\{(\Ra p_1),(p_2\Ra)\}$,
using only cuts on $p_1$ or $p_2$. Hence (3) implies (1).
\qed

\begin{cor}
If ${\G}$ is a coherent canonical system in $\lp$ then 
$\tup{\lp,\vD_{\G}}$ (or equivalently $\tup{\lp,\vd_{\G}}$) is an extended logic.
\qed
\end{cor}

\subsection{Strict Canonical Systems}

In \cite{AL10} {\em strict} single-conclusion canonical systems were investigated.
These systems are canonical systems,
in which derivations can only contain definite sequents.
Now we show that the results of \cite{AL10} about these systems 
can be derived from results of the present paper.
For this purpose, we concentrate on a smaller set of canonical systems, 
for which we are able to strengthen Corollary \ref{equivalences}.

\begin{defi}
\label{definite system}
A canonical system is called {\em definite} 
if its right-introduction rules have only definite clauses as premises,
and its left-introduction rules have only definite clauses as hard premises.
\end{defi}

\begin{exa}
Every canonical system in which the set of logical rules is a subset of the set of rules 
for $\su,\perp,\leadsto,\affirm$ (of Example \ref{canonical rules examples}) is definite.
\end{exa}

\begin{cor}
\label{equivalences definite}
The conditions below are equivalent for a 
definite canonical system ${\G}$:
\be
\item ${\G}$ is strongly consistent.
\item ${\G}$ is coherent.
\item ${\G}$ admits strong cut-elimination.
\item ${\G}$ admits cut-elimination.
\item ${\G}$ is consistent.
\ee
\end{cor}
\proof
(1),(2),(3) are equivalent by Corollary \ref{equivalences} for 
every canonical system.
(3) trivially implies (4). 
(4) implies (5), since in a canonical system
there is no way to derive $p_1\Ra p_2$ without using cuts.
Finally, a proof similar to that of Theorem 1 in \cite{AL10}, 
(or Theorem \ref{strong consistency -> coherence} of this paper) shows that 
(5) implies (2).
\qed

\begin{rem}
Strong cut-elimination and cut-elimination are not equivalent in the general case.
To see this, consider the system $\G$ given in Example \ref{G with circle}.
As explained there, a sequent $\g\Ra E$ can be proved in $\G$ from no assumptions
iff it is of the form $\g\Ra\psi$, where $\circ^n\psi\in\g$ for some $n\geq 0$.
It is easy to see that every sequent of this form can be proved without using cuts,
and so $\G$ admits cut-elimination.
However, $\G$ does not admit strong cut-elimination. For example, 
one must apply cut on $\circ p$ to derive the empty sequent from the sequent $p_1\Ra\ $.
\end{rem}

To derive results about {\em strict} canonical systems, we prove the following lemma.

\begin{lem}
\label{strict lemma}
Let ${\G}$ be a definite canonical system,
and let $\Ss \cup \{s\}$ be a set of definite sequents.
If there exists a proof $P$ of $s$ from $\Ss$ in $\G$,
then there also exists a proof $P'$ of $s$ from $\Ss$ 
in which every sequent is a definite sequent,
and every cut formula in $P'$ also serves as a cut-formula in $P$.
\end{lem}
\proof
It is easy to see that starting from definite assumptions,
the only way one can produce a negative sequent in a definite canonical system 
is by an application of a left-introduction rule of the form:
\[\derb{\{\g,\sigma(\Pi_i) \Ra \sigma(E_i)\}_{1 \leq i \leq m}}
 {\{\g,\sigma(\Sigma_i) \Ra \}_{1 \leq i \leq k}}
 {\g, \sigma(\dd(\ptil))\Ra}\]
Since $\G$ is definite, the sequent inferred in steps of this kind
cannot be used in the rest of the proof, unless right weakening is applied on a descendant of this sequent.
Applying the same weakening before steps of this kind will turn 
the sequent into a definite one, keeping the rest of the proof valid.
Finally, this modification does not affect the set of cut-formulas used in the proof.
\qed

Now define a new {\em strict} provability relation $\vd_{\G}^{seq_1}$ for definite canonical systems.
$\vd_{\G}^{seq_1}$ is defined as in Definition \ref{syntactic tcr}, except that
it  allows only {\em definite} sequents in proofs.
By Lemma \ref{strict lemma} it immediately follows 
that a definite system admits cut-elimination with respect to $\vd_{\G}^{seq_1}$, 
iff it admits cut-elimination with respect to $\vd_{\G}^{seq}$.
The same applies to strong cut-elimination and consistency.
Therefore for definite canonical systems,
Corollary \ref{equivalences definite} ensures that
coherence, cut-elimination, strong cut-elimination, and consistency\footnote{Note that strong consistency
is trivial in this case, since the empty sequent is not allowed to appear in derivations.}
are equivalent also with respect to $\vd_{\G}^{seq_1}$.

\section{Analycity and Decidability}

In this section we show that the 
semantics of ${\G}$-legal frames is analytic in the intuitive sense described 
in Remark \ref{analytic remark}.


\begin{thm}[Analycity]
\label{analycity}
Let $\U_1,\U_2$ be sets of formulas closed under subformulas, 
such that $\U_1\subset\U_2$.
Let ${\G}$ be a coherent canonical system for $\lp$.
The semantics of ${\G}$-legal frames is {\em analytic}
in the following sense: 
If ${\W_1=\tup{W,\leq,v_1}}$ is a ${\G}$-legal $\U_1$-semiframe, 
then $v_1$ can be extended to a function $v_2$ so that 
${\W_2=\tup{W, \leq,v_2}}$ is a ${\G}$-legal $\U_2$-semiframe.
\end{thm}
\proof
Similar to the proof of Theorem 6 from \cite{AL10}.
\qed

\begin{rem}
In particular, the last theorem shows that 
every ${\G}$-legal $\U$-semiframe, 
can be extended to a ${\G}$-legal $\lp$-frame.
\end{rem}

The following two theorems are consequences 
of Theorem \ref{analycity} and the
soundness and completeness theorems.

\begin{thm}[Conservativity]
\label{conservativity}
Let ${\bf G_1}$ be a coherent canonical system
in a language $\lp_1$, and let ${\bf G_2}$ be a 
coherent canonical  system in a language $\lp_2$.
Assume that $\lp_2$ is an extension of $\lp_1$
by some set of connectives, and that ${\bf G_2}$
is obtained from ${\bf G_1}$ by adding to the latter
canonical rules for connectives in $\lp_2-\lp_1$.
Then ${\bf G_2}$ is a conservative extension of ${\bf G_1}$
(i.e.: if all sequents in $\Ss\cup s$
are in $\lp_1$ then $\Ss\vd_{\bf G_1}^{seq} s$
iff $\Ss\vd_{\bf G_2}^{seq} s$).
\end{thm}
\proof
Suppose that ${\Ss}\not\vd_{\bf G_1}^{seq} s$. 
Then there is ${\bf G_1}$-legal model $\W$ of $\Ss$ which is not a model of $s$. 
Since the set of formulas of $\lp_1$ is a subset of the set of formulas of $\lp_2$ which is
closed under subformulas, 
Theorem \ref{analycity} implies that $\W$ can be extended to a ${\bf G_2}$-legal model 
of $\Ss$ which is not a model of $s$.
Hence ${\Ss}\not\vd_{\bf G_2}^{seq} s$.
\qed

\begin{thm}[Decidability]
\label{decidability} 
Let ${\G}$ be a coherent canonical system. 
Then ${\G}$ is strongly decidable:
Given a finite set $\Ss$ of sequents, and a sequent $s$,
it is decidable whether $\Ss\vd_{{\G}}^{seq}s$ or not.
\end{thm}
\proof
Let ${\U}$ be the set of subformulas in
$\Ss\cup\{s\}$. From Theorem \ref{analycity} and the proof
of Theorem \ref{key} it easily follows that
in order to decide whether ${\Ss\vd_{{\G}}^{seq}s}$
it suffices to check all triples of the form $\tup{W,\suq,v^\prime}$
where $W\suq 2^\U$ and 
${v^\prime: W \times \U  \to \{t,f\}}$,
and see if any of them is a ${\G}$-legal $\U$-semiframe
which is a model of $\Ss$ but not a model of $s$.
\qed

\begin{rem}
The last two theorems can also be proved directly 
from the cut-elimination theorem.
\end{rem}

Strong conservativity and strong decidability of $\vd_{\G}$ and $\vD_{\G}$ 
are easy corollaries of the previous theorems and the reductions given in 
Proposition \ref{reduction1} and Proposition \ref{reduction2}.

\section{Conclusions and Further Work}

Now we present our answer to the question from the introduction:
``what is a basic constructive connective?".

\smallskip
\begin{quote}
{\em A basic constructive connective is a connective 
defined by a set of rules in some coherent canonical system.}
\end{quote}

\smallskip\noindent
Theorem \ref{cut-elimination} ensures that the proof-theoretic criterion for constructivity, described in the
introduction, is met. Theorem \ref{conservativity}
ensures that a set of rules for some connective can indeed be seen as a definition of that connective, 
because it shows that in coherent canonical systems the same set of rules defines the same connective
regardless of the rules for the other connectives.


In Section \ref{semantic section}, the proof-theoretic characterization of basic constructive connectives 
was matched by a (non-deterministic) Kripke-style semantics. This semantics is modular, allowing to separate
the semantic effect of each derivation rule. However, we did not provide there
an independent semantic characterization
of (basic) constructive connectives. 
We leave this issue to a future work.
Another future goal is to extend our results to first-order logic, 
and identify constructive quantifiers as well
(for semi-classical quantifiers this was done in \cite{AZ08}).

\section*{Acknowledgements}
We are grateful to two anonymous referees for their helpful suggestions
and comments.
This research was supported by The Israel Science Foundation (grant no. 280-10).


\begin{thebibliography}{99}

\bibitem{av:9simple}
Avron, A.: Simple Consequence Relations. 
Information and Computation 92, 105--139 (1991).

\bibitem{Av93} 
Avron, A.: Gentzen-Type Systems, Resolution and Tableaux.
Journal of Automated Reasoning 10, 265--281 (1993).


\bibitem{AL05} 
Avron, A., Lev, I.: Non-deterministic Multiple-valued Structures.
Journal of Logic and Computation 15, 24-261 (2005). A partial
conference version in Gor\'{e}, R., Leitsch, A., Nipkow, T., (eds.):
Proceedings of  IJCAR 2001.
LNCS (LNAI), vol. 2083, pp. 529--544.
Springer, Heidelberg (2001).

\bibitem{AL10} Avron, A., Lahav, O.: Strict Canonical constructive systems.
In Blass, A., Dershowitz, N., Reisig, W. (eds.): 
Fields of Logic and Computation:
Essays Dedicated to Yuri Gurevich on the Occasion of His 70th Birthday,
75--94, 
Lecture Notes in Computer Science, volume 6300, Springer-Verlag, 2010.
A conference version in M. Giese, A. Waaler (eds.):
Proceedings of TABLEAUX 2009, 62--76, LNAI 5607, Springer (2009).

\bibitem{AZ08} Avron, A., Zamansky, A.: Canonical Gentzen-type calculi with (n,k)-ary quantifiers.
Logical Methods in Computer Science 4, 1--23 (2008).

\bibitem{belnap-tonk} 
Belnap, N. D.: Tonk, Plonk and Plink. Analysis 22, 130--134 (1962).

\bibitem{Bowen} 
Bowen, K. A.: An extension of the intuitionistic propositional calculus.
Indagationes Mathematicae 33, 287--294 (1971).

\bibitem{CT06} 
Ciabattoni, A.,  Terui, K.:
Towards a Semantic Characterization of Cut-Elimination.
Studia Logica  82, 95--119 (2006).

\bibitem{gentzen69}
Gentzen, G.: Investigations into Logical Deduction.
In: Szabo, M.E. (ed.) The Collected Works of Gerhard Gentzen,
pp. 68--131. North Holland, Amsterdam (1969).

\bibitem{gu09} 
Gurevich, Y., Neeman, I.: The Infon Logic: the Propositional Case.
To appear in ACM Transactions on Computation Logic 12 (2011).
An earlier version in 
Bulletin of European Association of Theoretical Computer Science,
number 98 (2009) 150--178.

\bibitem{Kr65} 
Kripke, S.: Semantical Analysis of Intuitionistic Logic I.
In: Crossly, J., Dummett, M. (eds.)
Formal Systems and Recursive Functions, pp. 92--129.
North-Holland, Amsterdam (1965).

\bibitem{McCullough}
McCullough, D.P.: Logical connectives for intuitionistic propositional logic.
Journal of Symbolic Logic 36(1), 15--20 (1971).

\bibitem{pr:60} 
Prior, A.N.: The Runabout Inference Ticket.
Analysis 21, 38--39 (1960).

\bibitem{handbook-sundholm}
Sundholm, G.: Proof theory and Meaning.
In: Gabbay, D.M., Guenthner, F. (eds.)
Handbook of Philosophical Logic, vol. 9, pp. 165--198 (2002).

\end{thebibliography}
\end{document}